\documentstyle[preprint,eqsecnum,aps]{revtex}
\def\beq{\begin{equation}}
\def\eeq{\end{equation}}
\begin{document}
\preprint{UW/PT-94-12}
\title{Long Range Forces in Quantum Gravity}

\author{Ivan J.\ Muzinich}
\address{Lauritsen Laboratory of High Energy Physics\\
California Institute of Technology\\
Pasadena, CA 91125 452-48\\
{\rm E-mail: muzinich@theory.caltech.edu}}

\author{Stamatis Vokos}
\address{Department of Physics, NK-15\\
University of Washington\\
Seattle, WA 98195\\
{\rm E-mail: vokos@phys.washington.edu}}
\date{December 1994}
\maketitle
\vspace*{-1cm}
\begin{abstract}
We calculate the leading quantum and semi-classical corrections to the
Newtonian potential energy of two widely separated static masses.
In this large-distance, static limit, the quantum
behaviour of the sources does not contribute to the quantum
corrections of the potential. These arise exclusively from the
propagation of massless degrees of freedom.
Our one-loop result is based on Modanese's
formulation and is in disagreement with Donoghue's
recent calculation. Also, we compare and contrast
the structural similarities of our approach to
scattering at ultra-high energy and large
impact parameter. We connect our approach to results from
string perturbation theory.

\end{abstract}

\widetext
\section{Introduction}

Quantum gravity is a subject beset with calculational difficulties, most
predominantly due to ultraviolet divergences that are incumbent in any
4-D theory with a coupling constant with dimensions of length.
These difficulties are well documented in the literature and
have not allowed a consistent formulation of the theory at short
distances \cite{rpf} . The hope is that superstring theories will
provide us with a robust definition of the theory in that regime.
However, as Donoghue \cite{donoghue} recently pointed out, one can
analyze quantized general relativity within the computational
framework of effective field theory.
He argued and showed that the unknown ultraviolet structure of the
theory is irrelevant for the purposes of obtaining well-defined
leading-order in $\hbar$ quantum corrections to some
quantities involving gravitational effects at large distances.
It is possible to formulate this program in a general way
armed with techniques based upon general analyticity properties and
gravitational Ward-Slavnov-Taylor identities. As we will discuss, this strategy
is also implicit in the analysis of the scattering problem at
ultra-high energy
and large impact parameter considered by a number of authors
\cite{sol,thooft,oth} in the past few years. The problem of long range
forces is primarily a consequence of soft virtual gravitons and
inherently of an infrared
nature. The infrared region with respect to radiation of soft
gravitons and the Block-Nordsieck \cite{bn} cancellation of infrared
singularities was communicated long ago by S.\ Weinberg \cite{weinberg} and
B.\ DeWitt \cite{witt}.

Donoghue employed this method to obtain the leading corrections to
the Newtonian potential. Our starting point is similar in spirit but
quite different in implementation.
We present and apply a general
formalism presented recently by Modanese \cite{mod} which we believe is
better suited for the study of the
simplest quantity in the long distance regime, namely the
static potential. We obtain a different result than Donoghue, and we
discuss the reasons for the discrepancy in detail. In particular,
we compute the quantum corrections in the gravitational or more
generally the zero-mass sector of the theory in a manner that can be
applied to either the static or ultra-relativistic problems. This is
in parallel with studies of non-Abelian gauge theories.
\cite{adm,fischler,feinberg,lsb}.  The approach of Modanese
is presented in terms of the Euclidean functional integral and
parallels the Wilson loop of gauge theory. The formalism is quite
general and allows a starting point for perturbative investigations of
the long distance regime of this work as well as non-perturbative
numerical simulations. On the other hand, the application to the
relativistic scattering domain has been communicated in the literature
under the names of relativistic eikonal, and semi-classical (WKB)
techniques by many authors \cite{lev,cheng,itz}. The two problems are very
closely related and a general result can be derived by functional
integral techniques
which can be characterized by the exponentiation of certain connected Green's
functions. The connected nature of the exponentiation of the Green's
functions separates the potential from its iterations in a clear
manner.

The actual work in such calculations is two fold. Firstly, one
identifies a formulation that allows a convenient isolation of the
effective theory in the relevant kinematic domains. In the static
limit the relevant effective theory is a 3-D, dimensionally reduced
quantum gravity, while in the ultrarelativistic domain, the effective theory is
2-D. Secondly, the isolation of contributions
that survive as inverse powers at large distances, i.e.\ as $R^{-n}$
for $n>1$ correspond to
singular contributions at small momentum transfer i.e.\ analytic
 in the cut $Q^2$ plane with branch cut singularities at zero
momentum transfer. The Fourier analysis of the Newtonian potential
($n=1$) is already singular at small momentum transfer with the well
known pole in $Q^2$. The specific power and nature of the
singularities depends upon the number of graviton interactions and
other details that will be elucidated in the following sections.

Contributions analytic at small momentum transfers (in the sense of
old S-matrix theory) correspond to forces of finite range of the form
$\exp -(R/L)$ due to the Riemann-Lebesgue Lemma,
and permit a Taylor series expansion at small
momentum transfer. It is tempting to say that modern superstring
theory will relate the range or the effective length parameter to the
string tension in a way that all such contributions can be rendered
finite. However, tempting such a correspondence is, its proof certainly
lies outside the scope of this investigation.
Nevertheless, we will expound on this idea using known one-loop
results from string theory in a later section of this
communication. The field theory limit with zero mass particles in
loops arises from string interactions (higher genus Riemann surfaces),
as required by unitarity. This limit is achieved at zero string
tension, with polynomial corrections with finite coefficients which
characterize the surviving string interactions at low energy. It is
only the effective field theoretic (or S-matrix) contributions with
zero mass intermediate states that reduce to non-polynomial
interactions at low energies.

The organization of this communication is as follows: In Section II,
the underlying technique pioneered by Feynman\cite{feynman} and developed
by DeWitt\cite{witt} and Mandelstam\cite{mandelstam} is briefly reviewed.
In this approach the metric is treated as a tensor field on a flat
space-time background in parallel with any other quantum field. Path
integral quantization
is used as a convenient framework to introduce external sources to
formulate the static potential (or scattering amplitude).
The recent work of Modanese is presented, as well as mention of the
older work of many authors on the high energy scattering. The relevant
dimensionally reduced effective field theories are identified. A
discussion of the results in terms of connected graphs of the
gravitational and more generally the zero mass sector is
presented. Section II also deals with the calculational aspects of
this work, namely the relevance of the different distance scales and
the isolation of the power behaved contributions at large
distances. Calculations in the ultra-relativistic regime, relevant to the
scattering of constituents at Planckian energies are reviewed but not
presented in any great detail, since these have been dealt with in the
past few years by a CERN group \cite{ama}.

Following the explicit calculations, a general anatomy of a proof is
presented in Section III to indicate how one would make a general
proof based upon the plausible scenario that the small distance
effects do not survive at large distances. Here recent calculations of
higher order (higher genus surfaces) contributions in string theory
are quite necessary.

Finally, we consider the pitfalls and limitations of such an
investigation and compare and contrast gravity with other gauge theories, where
long range forces play out in different ways. In abelian gauge
theories the finite fermion mass screens all contributions other than
the Coulomb potential between charged particles in the deconfined
phase. There are also non-trivial, inverse-power behaved van der Waals
forces between neutral objects resulting from effective
non-renormalizable interactions at low energies \cite{sucher}. While
the techniques in gravitational theory are similar to the analysis of
molecular forces, the physics is vastly different. In non-abelian
theories, confinement
and spontaneous symmetry breakdown are plausible dynamical mechanisms
that render long range inverse power behaved forces inoperable. An appendix
deals with some technical questions on the details of the
calculations.

\section{Formulation}
\subsection{Static Limit}

The general expression for the static potential energy for the
gravitational interactions of two static massive particles requires
some care and has been given recently by G.~Modanese\cite{mod}. His work
builds upon some expositions of other authors in both the static limit
and the ultra-relativistic limits. Standard conventions
will be followed with a flat space metric signature given by
$\eta_{\mu\nu}={\rm diag}(1,-1,-1,-1)$: we will work in $M_4$
(4-D Minkowski space) instead of Euclidean space, since we will be
interested in the perturbative definition of the theory with no
applications to numerical simulations, or issues of instability.
The Einstein-Hilbert action is given by
\beq
S_{\rm EH}=-{1\over 16\pi G}\int dx\, \sqrt{g(x)}\, R(x)\,,
\eeq
where the integral is over the full four dimensional space time,
$g(x)=-{\rm det}(g_{\mu\nu})$ and $R$ is the curvature scalar. It suffices to
consider the action of two constituents, since we are considering the two body
potential. The coupling of the sources to the metric is via
\beq
S_{\rm M}=-M_1 \int ds_1 - M_2\int ds_2\,,
\eeq
where $ds=(g_{\mu\nu}(x(\tau))\dot{x}^\mu \dot{x}^\nu)^{1\over 2}
d\tau$ is the line element along the trajectory of each object of
mass $M_1$ and $M_2$  respectively,  $\tau$ is
the proper time measured along the trajectory, and $\dot{x}$ is the
covariant four-velocity. The generating functional for the
gravitational interactions between two such massive particles is given
by the functional integral:
\beq
Z_{\rm M}={\int [Dg]\, \exp {i\over \hbar}(S_{\rm EH} + S_{\rm
M})\over \int [Dg]\, \exp {i\over \hbar}S_{\rm EH}}\,.
\eeq
The integration is over the metric $g$ with some suitable gauge fixing
for the perturbative definition of the Feynman diagrams of the
theory. We seek the potential energy of two static sources situated
at
\beq
x_1^\mu = (t_1,\vec{0})\qquad\qquad
x_2^\mu = (t_2,\vec{R})\,,\qquad\qquad -T/2 \leq t_i\leq T/2
\eeq
These trajectories correspond to two static particles with coordinate
separation $R$ and four-velocities $\dot{x}_1^\mu=\dot{x}_2^\mu =
\delta^\mu_0$, with no relative motion. The limitations of the
restriction to static sources
within the context of the full dynamics is discussed below.

In the perturbative definition of the functional integral Eq.~(2.3),
the metric is expanded around the flat space metric $\eta_{\mu\nu}$
in the usual fashion of the Feynman-DeWitt method of quantization
of general relativity, namely
\beq
g_{\mu\nu}(x)=\eta_{\mu\nu} + \kappa h_{\mu\nu}(x)\,,
\eeq
where $\kappa^2=32 \pi G$. Following Symanzik\cite{sym}, we shall
identify the lowest energy of the system (the gravitational potential
energy, in the static case at hand), with the limit
\beq
V(R)=\lim_{T\to \infty} {i\hbar\over T}\, \ln Z_{\rm M}\,,
\eeq
in view of the identification of $Z_{\rm M}$ with the vacuum-to-vacuum
amplitude in the presence of external sources, given by $S_{\rm M}$,
adiabatically switched on at $-T/2$ and off at $T/2$,
\beq
Z_{\rm M} = \langle 0|e^{-{iHT/
\hbar}}|0\rangle=\sum_{n}e^{-{iE_nT/\hbar}}|\langle 0|n\rangle|^2\,,
\eeq
for a very large time $T$. It is clear that Eq.~(2.6) above is
meaningless as it stands. We give it meaning by endowing the large Minkowski
time interval $T$ with a small negative phase, or by keeping track of the
$i\epsilon$'s in the perturbative propagators.
This particular definition of the ground state energy was first
advocated by Symanzik and proven in perturbation theory for the case
of a source coupled linearly to the quantum field.  It has been used
in many other field theory
contexts. In the application to the static potential problem, this
formulation clearly differentiates the static potential from its
iterations, the higher order terms in $T$ of Eqs. (2.6) and (2.7).
We will assume that this definition holds even in the case presence of sources
coupled to higher-order local monomials of the field.

If $g$ is expanded as in Eq.\ (2.5) in Eq.~(2.3), we obtain (to lowest order
in $\kappa$) the customary
result,
\beq
Z_{\rm M}=\langle \exp{-{i\kappa\over 2\hbar}\int dx\, h_{\mu\nu}(x)
T^{\mu\nu}(x)}\rangle,
\eeq
where $T_{\mu\nu}$
is the classical energy-momentum tensor of the source, and
averages with respect to the normalized gravitational functional
integral are defined in the usual way,
\beq
\langle{\cal O}\rangle={\int [Dh] {\cal O} \exp{{i\over \hbar}
 S_{\rm EH}}\over \int [Dh] \exp{{i\over \hbar} S_{\rm EH}}}\,.
\eeq
A trivial time dependence proportional to $\exp
-i(M_1+M_2)T/\hbar$ may
be absorbed into the normalization of the functional integral.

In the case of
a pair of static heavy
particles of masses $M_1$ and $M_2$ with trajectories given by
Eq.~(2.5), we have
\beq
S_{\rm M} = -M_1\int dt_1 \sqrt{1+\kappa h_{00}(x(t_1))}-M_2
\int dt_2 \sqrt{1+\kappa h_{00}(x(t_2))}\,,
\eeq
and
\beq
T^{\mu\nu}(x)=M_1 \delta^\mu_0\delta^\nu_0 \delta^{(3)}(x-x(\tau_1))+
M_2\delta^\mu_0\delta^\nu_0 \delta^{(3)}(x-x(\tau_2))\,.
\eeq

We close this subsection with some remarks. In general the definition
of the potential energy Eq.~(2.6) and Eq.~(2.9) appear to be frame-dependent
and path-dependent in any field theory including gauge theory. The
analogue in gauge theory, the Wilson loop, is gauge invariant but
path-dependent unless the field strength tensor vanishes
identically\footnote{It is possible to construct a
reparametrization-invariant``Wilson-loop like'' quantity (see
\cite{mod})}. While the analogous expression in general relativity
does not appear generally covariant (gauge invariant), the results
Eqs.\ (2.3) and (2.10) are still a fully covariant definition of the
interaction in the static limit much in the same way that the Coulomb
interaction is a necessary ingredient in the choice of Coulomb gauge
in quantum electrodynamics for a consistent gauge invariant
dynamics. Of course it will be necessary to consider kinetic energy
and the full dynamics to establish  the general covariance in a
completely convincing manner.

The particular choice of trajectories Eq.~(2.4) insures that the static
potential alone is being calculated, and that no terms involving the
kinetic energy or relativistic corrections are being included. At the
level of $(v/c)^4$ and $V\cdot\ (v/c)^2$ relativistic corrections are
bound to become important in the determination of motion in accord
with the post Newtonian
approximation  of classical general relativity, where both terms of
order $GM/Rc^2$ and $(v/c)^2$ are retained. The virial theorem of
classical mechanics relates kinetic energy $Mv^2$ to potential energy
$V$.
Therefore, kinetic energy will have to be included to order $(v/c)^4$
in order to have a consistent dynamics for the purpose of computing
the subsequent motion of this pair of gravitating objects.
In our work Eq.\ (2.6) and the incumbent static sourses
Eq.~(2.4) form a convenient starting point for the computation of the
static potential. The last
remark underscores the statement above concerning gauge invariance
(general covariance).

Notice that in the weak coupling limit Eq.\ (2.8), the lowest-order
energy is invariant under infinitesimal coordinate transformations
i.e.
\beq
h_{\mu\nu}(x)\to h_{\mu\nu}(x)+\partial_\mu \xi_\nu(x)+\partial_\nu
\xi_\mu(x)\,,
\eeq
for a conserved energy momentum tensor in Eq.\ (2.8). The complete
static problem also has a partial covariance with respect to time
independent coordinate transformations that are in the little group of
$g_{00}(x_i(t))$, where $x_i(t)$ are the trajectories, Eq.~(2.4).

It is also worth mentioning that neither Eqs.\ (2.6) or (2.8) is a loop
integral, since the notion of oppositely charged sources does not
exist for a tensor theory of gravity. This property
is essential in obtaining the Wilson loop for the static energy in
gauge theories. One may also consider the possible physical meaning
of a gravitational loop functional or holonomy,
\beq
{\cal U}(C) = -4 + {\rm Tr} {\cal P}\langle
\exp{i\oint_C \Gamma_\mu dx^\mu}\rangle\,,
\eeq
where one views the connection coefficients
$\Gamma^{\nu}_{\mu\lambda}$ as matrix elements of a matrix
$\Gamma_\mu$. This issue was investigated by Modanese
\cite{modanese2} . He showed that the expectation value of the loop,
calculated perturbatively around a flat background,
does not give rise to long-range interactions, and so cannot
be interpreted as representing a gravitational potential.
The physical interpretation of such a result can be understood in
terms of the functional integral. To leading order in $\hbar$,
the vanishing of the holonomy implies that the weak field
configurations which contribute to the functional integral have zero
curvature (to that order). An intuitive way to understand this
discrepancy between gravity and gauge theories is by noting that the
coupling of the vector field to matter is in terms of $\int d^4 x
A_\mu j^\mu$. For a pair of oppositely charged static sources, this
coupling reduces to the Wilson-loop exponent. For gravity, on the
other hand, the coupling to a source is non-linear in the metric (see
Eq.~(2.10)). Even the perturbative expansion
involves a coupling of $h_{\mu\nu}$ with the energy-momentum tensor
$T_{\mu\nu}$ (see Eq.~(2.8)). The affine connection coefficients simply
do not arise in such an expansion.


We proceed next to the calculations of the first-order corrections in
$G$, both semi-classical and quantum mechanical,
to the classical Newtonian potential energy,
using the formalism developed in the previous sections. The choice of
trajectories in Eq.\ 2.4 ensures that the static limit is retained as
opposed to calculating any effects associated with relativistic
corrections. In the following the harmonic or de Donder gauge
is maintained,
\beq
\partial_\mu(h^{\mu\nu}-{1\over 2}
\eta^{\mu\nu}h^{\lambda}_{\lambda})=0 \,.
\eeq
In the harmonic gauge the graviton propagator is simply
\beq
D_{\mu\nu\rho\sigma}(x)\equiv P_{\mu\nu\rho\sigma} \Delta(x)=
{1\over 2}(\eta_{\mu\rho}\eta_{\nu\sigma}+
\eta_{\mu\sigma}\eta_{\nu\rho}-\eta_{\mu\nu}\eta_{\rho\sigma})\Delta(x)\,,
\eeq
where
\beq
\Delta(x)=-{1\over 4\pi^2}{1\over x^2-i\epsilon}\,,
\eeq
is the free massless propagator. The choice of gauge restricts
the form of Green's
functions and the perturbative Feynman rules as expected. An appendix
deals with these and related issues in some detail. It is also assumed
that the cosmological constant is zero. This is very important, a
non-zero cosmological constant is an obstruction to the long distance
behavior of Green's functions and their incumbent Ward-Slavnov-Taylor
identities.

Next we apply the strategy given in Eqs. (2.3) and (2.10) to calculate the
potential to order $G^2$. The basic strategy is to expand the matter
action Eq.\ (2.10) to order $G$ ($\kappa^2$),
\begin{eqnarray}
S_{\rm M} &=& -(M_1+M_2)T+ \sum_{i=1}^2 \left(-{\kappa M_i\over 2} \int dt_i
h_{00}(x(t_i))+{\kappa^2M_i\over 8} \int dt_i h_{00}
(x(t_i))^2+\ldots\right)\\
&\equiv&  -(M_1+M_2)T+\sum_{i=1}^2 \left(J_i \int dt_i h_{00}(x(t_i))
+ K_i \int dt_i h_{00}(x(t_i))^2+\ldots\right)
\end{eqnarray}
and expand the partition function in powers of the sources $J_i$ and $K_i$.
Next connected graphs in the
gravitational sector are retained. The graphs are connected to sources
which are vertices for graviton lines. The complete set of graphs
contributing to the static potential to order $\kappa^4$ ($G^2$) is
shown in Figs.~ 1 to 3.
The time coordinates of the sources are to be integrated from $-T/2$
to $T/2$ and
we are left with an effective 3-D theory. The sources of mass $M_1$
and $M_2$ are
coupled to the gravitational sector at positions $\vec{0}$ and
$\vec{R}$.
The potential or
static energy can be read off from Eq.\ (2.6) in the large $T$ limit.

The connected nature of the Green's functions of $h_{00}(x)$ and local
monomials thereof is extremely important and differentiates the
potential from its iterations, which are terms of higher order in $T$
in the expansion of the exponential in Eqs.\ (2.17), (2.18).

The contribution to lowest order in $G$  can be computed from the
single graviton exchange graph in
Fig.\ 1 \cite{hamber}. The result is straightforward and familiar
\beq
f_1  =  \left(-{i\kappa M_1\over 2}\right)\left(-{i\kappa
M_2 \over 2}\right) \left(-{1\over 8 \pi^2}\right)
\int_{-T/2}^{T/2}
dt_1 dt_2\, {1\over (t_1-t_2)^2 -R^2 -i\epsilon}\\
\eeq
Expressing the integrand as a contour integral
and integrating gives (in the limit of large $T$)
\beq
f_1 = i T {G M_1M_2\over R}\,,
\eeq
which, in view of Eq.\ (2.6),  yields the Newtonian potential energy
\beq
V(R)=-{GM_1M_2\over R}\,.
\eeq
The corrections to order $G^2$ are of two classes: classical general
relativistic (independent of $\hbar$) and quantum mechanical of order
$\hbar$.
They are enumerated in Fig.\ 2 for the
semi-classical contributions and Fig.\ 3 for the quantum
contributions. It is easy to understand that a perturbative expansion
of the functional integral in gravity gives us classical
contributions, both formally and intuitively. The formal reason is
that the expansion includes tree graphs connected to an arbitrarily
high number of external classical sources. Boulware and Deser\cite{db}
showed that the tree graphs reproduce classical general relativity
(see also \cite{bb,duffsch}). The intuitive reason is that in gravity
there are two dimensionless parameters that lend themselves to a
perturbative expansion, one which involves $\hbar$ and is the ratio of
the Planck length with the separation of constituents, and one which is
independent of $\hbar$ and is the ratio of the Schwarzchild radius of
a mass with the separation of constituents.

The quantum contributions involve closed loops of
gravitons and are of order $\hbar$.
Fig.~3a is the vacuum polarization and receives
contribution from the Fadeev-Popov-Feynman ghost which is a vector
particle in gravity; this contribution is
well known in the literature and has been calculated long ago.

The calculations are discussed in detail in an appendix.
We note the results here. The corrections to the Newtonian potential
are denoted by $\delta V$ and the results are as follows,
\begin{eqnarray}
\delta V_{\rm Cl}(R)&=& - {GM_1M_2\over R}\left({G(M_1+M_2)\over
Rc^2}(-1+{1\over 2})\right)\\
                    &=& - {GM_1M_2\over R}\left(-{G(M_1+M_2)\over 2Rc^2}\right)
\end{eqnarray}
from the graphs of Figs.\ 2a and 2b respectively.

The quantum correction is of the form
\begin{eqnarray}
\delta V_Q (R) &=&  - {GM_1M_2\over R} \left({G\hbar\over\pi R^2c^3} ({43\over
30}+{1\over 4}-{5\over 6})\right)\\
&=& - {GM_1M_2\over R} \, {17 \over 20\pi}{G\hbar\over R^2c^3}
\end{eqnarray}
The detailed numbers come from
Figs.\ 3a, 3b, and 3c respectively. In
all of these contributions, there is no difficulty in extracting the
large time behavior in the leading static limit for large masses neglecting
recoil in Eq.\ (2.6).
Putting everything together gives to order $G^2$,
\beq
V(R) = -{GM_1M_2\over R}\left(1 -{G(M_1+M_2)\over 2Rc^2} +
{17 G\hbar\over 20\pi R^2c^3}\right)\,.
\eeq

All other contributions from skeleton graphs to this order are of short
range and have support in a region where we do not believe the
calculation in any case.

We close this section with some remarks about the regime of validity
of our results.
\begin{itemize}
\item There is a hierarchy of distance scales in the problem, namely
the Compton wavelength of the sources $h/M_ic$, the Planck length
$(G\hbar/c^3)^{1/2}$,
and the Schwarzchild radii of the sources, $2GM_i/c^2$.
These scales are clearly not
independent, since the Planck length is the geometric mean of the
other two length scales. The corrections to the potential are to be
trusted for distances large compared to all of these distance
scales. This follows from the explicit expression of the corrections,
but most importantly it is implicit in our restriction of staticity.
Let us explain this more fully with a familiar Feyman diagramatic approach.
Consider a particle interacting with a massive center through the
exchange of a scalar particle (for convenience, not necessity). If the
momentum of the exchanged scalar is $q$ and the initial and final
momenta of the scattered particle are $p$ and $p'$, then the
scattering amplitude is proportional to the sum of the direct and
crossed propagators, namely
\beq
{i\over (p+q)^2 -m^2+i\epsilon}+{i\over (p'-q)^2-m^2+i\epsilon}\,.
\eeq
In the limit of large mass and assuming on-shell scattering,
$p_0\approx p'_0 \approx m$, and the above sum becomes
\beq
{i\over 2 m q_0 +i\epsilon} +{i\over -2 m q_0 +i\epsilon}={2
\epsilon\over \epsilon^2 + 4 m^2q_0^2} \propto \delta(q_0)\,.
\eeq
Numerator factors of $q_0$ have been neglected in addition to the
customary linearization of propagators (Recall that spin 1 and spin
two exchang will involve polynomials in $p+p'$. In this limit, one can
sum {\em all} the possible
 exchange
contributions to scattering, in view of the following
identity\cite{am}
\begin{eqnarray*}
2\pi \delta({\textstyle\sum} q_{i0}) &\times& \sum_{{\rm
perms}(q_{10}, \cdots,k_{n0})}
{i\over q_{10}+i\epsilon}\times {i\over q_{10}+q_{20}+i\epsilon}\times
\cdots\\
&\cdots& \times {i\over q_{10}+\cdots+q_{n-1,0}+i\epsilon} = (2 \pi)^n
\delta (q_{10})\times\cdots\times \delta(q_{n0})\,,
\end{eqnarray*}
where the delta function on the left hand-side is the overall energy
conserving $\delta$-function
$\delta({\textstyle\sum}q_{i0}+p_0-p'_0)\approx
\delta({\textstyle\sum} q_{i0})$. This identity shows that the sum
over the {\em dynamical} exchanges is equal to a single graph in which
the scattered particle field is replaced by a static pointlike
classical source $j(\vec{x})=\delta(\vec{x})= {\rm
F.T.}[2\pi\delta(q_0)]$.  This argument shows that we can neglect the
dynamical, quantum behavior of the field which undergoes the
scattering, provided that we are at a distance larger than {\em its} Compton
wavelength and larger than distance scales where non-perturbative
effects become important. The contribution of powers of $q_0$ in the
numerator will enter at order $q_0^2$ or higher in accordance with
expectations on the low energy behavior of quantum electrodynamics and
quantum gravity. It is also clear that these tecnniques can be
generalized to establish a result that allows the factorization and
the incumbent exponentiation for any finite correlation of exchanged
quanta coupled to the source. This argument, which is not presented in
detail here, is useful in establishing the general result in terms of
connected Green's functions Eqs. (2.10, 2.11, 2.17, 2.18) for higher
polynomials in $h_{00}$.

The above argument implies that the corrections to the potential are valid for
distances larger than the Compton wavelength of the source, if the
mass of the source is less than the Planck mass ($10^{-5}g$) (with
recoil corrections of order $q^2/M_i$). This includes all of the
elementary particles of the standard model including the hypothetical
vector bosons that mediate the baryon number non-conserving
interactions of Grand Unified Theories and other massive particles of
broken supersymmetric and technicolor theories. For
sources with macroscopic masses, on the other hand, the distance
should be greater than the Schwarschild radius of the source.
\item While the main results are similar in spirit to the work
of Donoghue \cite{donoghue}, the details and results differ. In our work the
coupling to static sources Eqs. (2.6) and (2.10) is maintained
throughout the entire calculation. In the calculation of Donoghue, the
scattering amplitude with recoil and relativistic corrections in the
coupling to matter are retained. The potential was defined as the
non-relativistic limit of the one particle reducible graphs of the
$t$-channel, a definition substantially different than
the one presented here. An important check on the classical correction
to the Newtonian
potential obtained here is the complete agreement with the post-Newtonian
potential energy of a test particle of mass $M_2$ in an external field
produced by $M_1$, in the limit $M_1\gg M_2$ and where $M_2$ has
zero velocity, namely $V=\Phi + 1/2 \Phi^2$, where
$\Phi$ is the Newtonian potential\cite{weinbergsbook}.
Our result for the potential energy
is also consistent with the expansion of the $00$ component of the
metric in a Schwarzschild geometry in isotropic coordinates,
\begin{eqnarray*}
g_{00}=1-{2 G M\over R}\left(1-{GM\over R}\right)\ldots\,,\nonumber
\end{eqnarray*}
with the proper post-Newtonian identification $g_{00}=1+2 \Phi + 2 \Phi^2$.
\item The quantum corrections in Eq.~(2.26) have a definite sign. The
physical origin of this sign is due to positivity of the Hilbert space
metric over physical states and is discussed in more detail in the
appendix. Notice that the Newtonian limit is approached from above and
the attractive interaction increases as the distance decreases. This
is in agreement with the analogous situation in quantum
electrodynamics with the form of the Uehling potential \cite{bjdr}.

\item A final word of caution is in order. It is only at the one-loop
level that the finite long distance behavior can be separated from the
divergent small distance counterterms in an unambiguous way,
at least with the field theory techniques presented so far.
Non-leading long distance quantum corrections at two loops and higher
will receive
contributions from divergent higher dimension polynomial counterterms.
In order to
completely understand the quantum contributions to long range forces
in higher orders, one would
need an improved definition of the theory. String theory is perhaps a
suitable candidate. We return to this question in section III.
\end{itemize}

\subsection{Ultra-Relativistic Limit}

The study of the elastic forward scattering of two scalar
particles at high
energies greater than the Planck mass, $s\gg M_{\rm
Pl}^2\gg t$, with $s=(p_1+p_2)^2$ and $t=(p_1-p_2)^2$,
has received attention in recent years because of its relation to
superstring theory.

One of the authors and M.~Soldate  investigated
the problem \cite{sol} because of the unitarity difficulty due to the
energy growth of the scattering amplitude in any theory of gravity in
near the forward scattering. 't Hooft and
others\cite{thooft,oth} were
interested in the problem because of its possible relation to black
hole dynamics, to string theory, and to other field theory
contexts. The scattering can be treated nicely by semi-classical
methods which go under the name of eikonal approximation or WKB and
are well developed in the literature. The basic strategy is to retain
the leading terms in the coupling of the external particles to
gravitons at high energies and linearize the propagators. Probably the
most lucid exposition again comes functional integral techniques which
we sketch briefly below. The four-point Green's function for the
scattering of four massive scalar particles, interacting only gravitationally,
is given by the time-ordered expression
\beq
\langle 0|T(\phi(x_1)\phi(x_2)\phi(x_3)\phi(0)|0\rangle=\int [Dh]
e^{{i\over \hbar}
S_{\rm EH}} G(x_1,x_3;g)G(x_2,0;g) + {\rm perms}\,,
\eeq
neglecting scalar loops, as they do not contribute to power behaved
long range contributions.
The propagators satisfy in
the massless limit (it suffices to retain the massless limit as an
approximation  in the
ultra-relativistic approximation),
\beq
-\Box G(x,x';g)={1\over \sqrt{g}}\delta^{(4)}(x-x')\,,
\eeq
where $\Box=g^{-1/2} \partial_\mu (g^{1/2}
g^{\mu\nu}\partial_\nu)$ is the generally covariant Laplacian.
If the metric is expanded about flat spacetime
and the leading terms at high energies are retained, one obtains
\beq
\left(-\eta^{\mu\nu}\partial_\mu\partial_\nu +
\tilde{V}(x)\right)G(x,x')=\delta^{(4)} (x-x')\,.
\eeq
In the leading approximation at high energy, the problem then reduces
to potential theory. The potential is:
\beq
\tilde{V}(x) = \kappa E^2 h_{--}(x)\,,
\eeq
for the right mover and ($-\to +$) for the left mover, where the
light-cone coordinates are defined as usual by $x^\pm = z\pm t$
(see in this context the papers by Kabat and Ortiz \cite{ko} and Verlinde
and Verlinde \cite{vv}). One passes to the amputated proper one-particle
propagator through the familiar scattering identities
$G=G_0+G_0\Sigma G_0=G_0+G_0 \tilde{V}G$,
 where $G_0$ is the free one-particle propagator.
Upon transformation to momentum space,
one obtains
\beq
\langle p'|\Sigma|p\rangle=\langle p'|\tilde{V}|\Psi_p\rangle=\int
dx\, e^{-i p'\cdot
x/\hbar}\tilde{V}(x) \Psi_p(x)\,,
\eeq
where $\Psi_p(x)$ is the scattering solution which satisfies $\Psi =
\Psi_0 + G_0\tilde{V} \Psi$ with appropriate
scattering boundary conditions. Its WKB scattering solution is
\beq
\Psi_p(x)= \exp {i\over \hbar}\left\{ p\cdot x - {1\over
2E}\int_{-\infty}^\tau \tilde{V}(x(\tau')) d\tau'\right\}\,,
\eeq
where $\tilde{V}(x)$ is given by Eq.~(2.32), and the trajectories
of the scattered constituents are given in light cone coordinates
$(+, {\rm tr}, -)$ by
\beq
x_1=(0,\vec{b},x^-)\,,\qquad\qquad x_2=(x^+,\vec{0},0)\,.
\eeq
In Eq.\ (2.34), $\tau$ is a coordinate that parameterizes the
trajectory, $\tau = x^+(x^-)$ for right(left) movers, and
$\vec{b}$ is a 2-vector representing the impact parameter. The on-shell
scattering amplitude results from amputation of the Green's function,
through use of Eqs. (2.29), (2.33), and (2.34),
\beq
{\cal A}(E,\vec{b})= 2 i E^2 \langle T\exp -{{i\over\hbar}}{\kappa\over 2}\int
h_{\mu\nu}T^{\mu\nu}\rangle\,,
\eeq
very much in parallel with Eq.~(2.8).   The energy momentum tensor in
Eq.~(2.36) now represents sources for the scattering of two
relativistic particles in the so-called shock wave kinematics,
\beq
T_{\mu\nu}= E\, \delta^{(3)}(x-x_1(\tau))\delta_\mu^+\delta_\nu^+ +
 E\, \delta^{(3)}(x-x_2(\tau))\delta_\mu^-\delta_\nu^-\,.
\eeq
The space-time coordinates in Eq.~(2.29) have been simplified by the
use of translational invariance and the form of the trajectories,
Eq.~(2.35) relevant to the scattering kinematics. The first term in
Eq.~(2.37) represents the right mover at impact parameter $\vec{b}$ and the
second term represents the left mover at the origin in impact
parameter space.

A familiar form for the scattering amplitude is derived from
Eq.~(2.36) after quadratic functional integration to order $G$ $(\kappa^2)$
of the gravitational action. Upon Fourier transformation to the transverse
momentum representation, one obtains in the relativistic limit, the
familiar eikonal formula with ${\cal A}= \exp i\Delta$,
\beq
{\cal M}(E,\vec{q}_{\rm tr})= 2 i E^2\int d\vec{q}_{\rm tr}\, \exp
i\vec{q}_{\rm
tr}\cdot \vec{b}\left(\exp i\Delta -1\right)\,.
\eeq
Equation (2.38) is the celebrated eikonal approximation corresponding
to the functional average in Eq.~(2.36); In lowest order the
contribution to the eikonal is given by the two dimensional graviton
propagator i.e. $\Delta\approx (-2 E^2 G)\ln \lambda b$, which
integrates to a form obtained in many references
\cite{sol,thooft,oth},
\beq
{\cal M}= 8\pi G {s^2\over t}\left\{\left({\lambda^2\over
t}\right)^{-i\alpha} {\Gamma(1-i\alpha)\over
\Gamma(1+i\alpha)}\right\}
\eeq
where $\alpha$ is the Regge trajectory for the Newtonian potential
$(\alpha={\rm O}(G{s\over \hbar}))$ as $s\to\infty)$,  $\lambda$ is an
infrared regulator for the 2-D effective massless propagator, and $s$
and $t$ are the usual Mandelstam variables. Eq.~(2.39) is valid for
large $s$ and, near forward scattering, at small $t$. The real term in
Eq.~(2.39) is the purely classical scattering, while the pure phase
(bracketed expression) is a contribution quantum mechanical in
origin. The latter term corrects the ill-behaved growth of partial
wave amplitudes at high $s$ and their incumbent violation of unitarity
\cite{sol}.

The general results Eqs. (2.36), (2.38), and (2.39)) have the desirable
properties and features:
\begin{itemize}
\item Unitarity in impact parameter space is retained at high values of
$s$. Unitarity
follows from reality of the exponential in Eq.~(2.40) for elastic
scattering and the positivity of its imaginary part for inelastic
scattering for radiation of soft gravitons. See in this context the
analysis of the CERN group in \cite{ama}. They establish a systematic
expansion for the analogue of the static semi-classical results in
$G^2s/b^2$.
\item The reader will see the close relation between the two contexts,
ultra-relativistic and static. The ultra-relativistic theory is simply
a 2-D image of the static 3-D effective theory.
\item The reader will also recall that Eqs. (2.36), (2.38), and (2.39)
together with the techniques used in their derivation were the basis
of phenomenological models of high energy hadron scattering at least
two decades ago \cite{cheng}. Such techniques produced a geometric picture of
high energy diffraction scattering of hadrons.
\end{itemize}

\section{Relation to String Theory and Discussion}

The detailed calculations of the quantum corrections for example are
presented only at the one-loop level, O$(\hbar)$. It is obvious that in
higher orders there will be difficulties. For example, divergent
subgraphs by power counting will necessitate higher dimension polynomial
valued counter terms coming from operators of dimension four and
greater such as $R^2$, where
$R$ is the Riemann tensor or any of its contractions. These divergent
counterterms can then propagate into the skeletons in calculations
analogous to those of Section III. It becomes impossible to proceed
further in a theory that is divergent and non-renormalizable such as
quantum gravity. Clearly some new definition of the theory is
necessary.

The only well developed strategy is superstring theory in one of its
various forms. We leave aside the vagaries of reconciling string
theory to the phenomenology of particle physics and dwell on the
question of a consistent theory of quantum gravity. Fortunately, there
are one-loop calculations of on-shell scattering amplitudes in the
literature \cite{mon,cap}. There are even calculations beyond one loop that
focus on unitarity which relate the higher loop graphs to tree graphs.
 The calculations are technical and involve the moduli space of
higher genus Riemann surfaces, the general name given to certain complex
manifolds that appear in string perturbation theory. The calculation of the
one-loop amplitude by Montag and Weisberger\cite{mon} is perhaps the most
relevant, and we will quote and interpret their results for our
purposes.

In string theory another scale emerges; namely the Regge slope or
string tension $\alpha'$ which plays two roles: It sets the scale of the
coupling to and the mass scale for the higher mass states in the
string spectrum of states. The field theory limit is then achieved by
taking the $\alpha'\to 0$ limit.

Higher-order graphs in string theory (of which there are fewer due to
duality, {\em e.g.}\ one at the one loop level)
involve integration over complex
coordinates in contrast to the integration over momenta of the
corresponding graphs in field theory.

To order $\kappa^2$, the structure of the tree plus one-loop, on-shell,
 invariant scattering amplitude for massless string constituents {\em
i.e.}\ dilatons
or gravitons is
\beq
{\cal M}(s,t,u)=K [ {\cal A}_0(s,t,u) + c {\cal A}_1(\alpha';s,t,u)]\,,
\eeq
where $K$ is a kinematical factor to ensure gauge invariance and proper
mass-shell behavior, and $c$ is a dimensionless constant. In Eq.\ (3.1),
${\cal A}_0$ is the typical Veneziano-type amplitude,
\beq
{\cal A}_0(s,t,u)={1\over stu}\, {\Gamma(1-{\alpha'\over
4}s)\Gamma(1-{\alpha'\over 4}t)\Gamma(1-{\alpha'\over 4}u)\over
\Gamma(1+{\alpha'\over 4}s) \Gamma(1+{\alpha'\over
4}t)\Gamma(1+{\alpha'\over 4}u)}\,,
\eeq
while ${\cal A}_1(\alpha';s,t,u)$ is the one-loop string amplitude,
the form of which is not essential for our purposes (it is given by an
integral of certain world-sheet
vertex operators over the moduli space of a torus). It suffices to
note that
in the low energy limit, the one-loop amplitude
can be expanded in powers of $\alpha'$ and the field theory limit is recovered
in the limit $\alpha'\to 0$, as long as $|\alpha' s|$, $|\alpha' t|$,
$|\alpha' u| \ll 1$. The expansion is non-trivial due to subtleties
in the region of integration in the moduli space
which ensure a finite amplitude in the physical region of $s,t,u$.
Nevertheless, when the dust settles, the result that emerges can
be written as an expansion, where the first non-trivial term is
$O(\alpha'^2)$ due to symmetric integration in moduli space is
\beq
{\cal A}_1(\alpha';s,t,u)={\cal A}_1^{(0)}(s,t,u)+\alpha'^2{\cal
A}_1^{(2)}(s,t,u)+\ldots\,,
\eeq
where ${\cal A}^{(0)}$ and ${\cal A}^{(2)}$
are the first two terms in the expansion in $\alpha'$. The first term
${\cal A}^{(0)}$ is
the zero slope or field theory limit, and the next term is a local
polynomial in the momenta at low energy, i.e.\
${\cal A}^{(2)}={\rm O}(s^2,t^2,u^2)$.
The field theory amplitude has a representation in terms of a
Feynman parameter integral,
\beq
{\cal A}_1^{(0)}=\int\prod_i d\beta_i\, \delta\left(1-\sum_j
\beta_j\right)\phi(s,t,\beta)^{(10-d)/2} + {\rm sym.}\,,
\eeq
where $\phi(s,t,\beta)=\beta_1\beta_2 s-\beta_3\beta_4 t$ is the usual
polynomial in momentum invariants, $d$ is the number
of compactified dimensions ($d=6$), and there is to be a symmetrization with
respect to the interchanges $s\to t$, $s\to u$, etc. At low energy the massive
states are completely decoupled and the field theory contributions  completely
dominate the amplitude.
Hence it is seen that the field theory
contributions overshadow the string corrections at low energies (large
distances), as expected also from the effective action for the
massless modes of the string (however, for a possible low-energy window on
string-scale physics see \cite{damour}).

Although the actual calculation presented here on the first
non-trivial loop correction is clearly suggestive, we expect that higher order
string calculations will support the general result of a
gauge-invariant amplitude with polynomial corrections to the field theory
result at low energies. At small distances, and $s$, $t$, and $u$ large and
comparable, the situation is quite different and string considerations
are important and essential \cite{men1,men2}.

We close this section with some further remarks concerning the
relation of our work on the long range forces in quantum gravity with
some related issues in gauge theories. Namely, dynamical mechanisms of
spontaneous symmetry breaking and mass generation render the Coulomb
potential to be the only leading long range force in electrodynamics
between charged objects. The finite fermion mass shields all long
range forces at the quantum level in such theories. In non-abelian
gauge theories such as QCD, confinement presumably renders all long
range forces inoperable, once the hadron mass spectrum is achieved in
a realistic calculational framework.

\acknowledgments
This work was supported by DOE grant DE-FG06-91ER40614. We are
indebted to David Boulware and Peter Arnold for uncountably many useful
conversations, to Patrick Huet and Larry Yaffe for
several discussions and to Sangyeong Jeon for teaching us tricks with
$i\epsilon$'s. S.V.\ thanks Lowell Brown for bringing the
problem to his attention and for many constructive suggestions.
Some tensor manipulations were carried out with MathTensor. The
figures were produced with Peter Arnold's Feynman diagram drawing program.
One of the authors, IJM, wishes to thank John Schwarz for hospitality
at the high energy theory group at Caltech where the final stages of
this manuscript were completed. He also thanks Stan Brodsky, Michael
Peskin, and other members of the SLAC theory group for stimulating
discussions during a seminar given there.

\appendix
\section{}

In this appendix we present some of the details of the
calculations. The Feynman rules in de Donder gauge have been presented
in the literature.  We will present what is sufficient for the
calculations at hand. The expansion of the metric Eqs. (2.5) and (2.10)
give us all of the coupling to the matter sources needed. The one- and
two-$h_{00}(x)$ graviton vertices that couple to the matter sources are simply
$(-\kappa iM_j/2)$ and $(i \kappa^2 M_j/4)$, respectively.
Furthermore, we need the triple graviton vertex. Actually, since we
require the three-$h_{00}$ Green's function, we only
need some special projections of the full triple vertex.
One can show that the momentum space transform of the configuration
space three-point function involving only $h_{00}$'s is given by the
projection
\begin{eqnarray}
V_3=V_{000000}&-&{1\over 2}(V_{0000}\,^\mu\,_\mu+V_{00}\,^\mu\,_\mu\,_{00}+
V^\mu\,_\mu\,_{0000})\nonumber\\&+&{1\over 4}(V_{00}\,^\mu\,_\mu\,^\nu\,_\nu +
V^\mu\,_\mu\,^\nu\,_\nu\,_{00}+V^\mu\,_\mu\,_{00}\,^\nu\,_\nu)\,
-{1\over 8}V^\mu\,_\mu\,^\nu\,_\nu\,^\lambda\,_\lambda\,,
\end{eqnarray}
where $V_{\kappa\lambda\mu\nu\rho\sigma}(p,q,-(p+q))$ is the full
triple graviton vertex in momentum space. The projected vertex is
\beq
V_3(p,q)=-{i\kappa\over 4}
\left(p^2+q^2+(p+q)^2-2(p_0q_0-(p_0+q_0)^2)\right)\,,
\eeq
These vertices are sufficient to compute the static potential to order
$G^2$. All contributions are relatively straightforward to compute except
for the vacuum polarization of Fig.\ 3. All computations are performed
in the harmonic
gauge of Eq.~(2.14).
The contribution to the connected Green's function of Eqs. (2.17) and (2.18)
from
the graph in Fig.\ 2a is
\beq
f_{2a}={i \kappa^3\over 16} M_1M_2(M_1+M_2)\int dt_1dt_2dt_3
G(t_1,\vec{0};t_2,\vec{0};t_3,\vec{R})\,.
\eeq
where $G(x_1,x_2,x_3)$ is the three-point function of the temporal
gravitons. Since all three gravitons are connected to static sources,
the $0$ components of their momenta vanish, and we obtain
that the three-point function in this case is the product of two
propagators. The time integrals are trivial delta functions of $p_0$
and $q_0$, times another delta function $2\pi \delta(0)=T$. Two of the
three contributions to the final Fourier integral which transforms
back to configuration space are divergent. However the divergence is a
ultra-local divergence, which is irrelevant to the long-distance
effects. The finite contribution is
\begin{eqnarray}
f_{2a}&=&{i \kappa^4\over 64} M_1M_2(M_1+M_2)T \left(i\int
{d^3\vec{p}\over (2\pi)^3} e^{-i \vec{p}\cdot \vec{R}}{1\over
\vec{p}^2}\right)^2\\
&=&-iT\, {GM_1M_2\over R}\, {G(M_1+M_2)\over R}\,.
\end{eqnarray}

Fig.\ 2b contributes
\beq
f_{2b}={1\over 2}\left(-{i \kappa\over 2}\right)^2
\left({i\kappa^2\over 4}\right) M_1M_2(M_1+M_2)\int dt dt_1 dt_2
\Delta(t-t_1,\vec{R}) \Delta(t-t_2,\vec{R})\,.
\eeq
A straightforward evaluation of the integral in Eq.\ (A.6) yields the
result
\beq
Re\, \int_{-{T\over 2}}^{{T\over 2}} dt dt_1 dt_2 {1\over
(t-t_1)^2-R^2-i\epsilon}{1\over (t-t_2)^2-R^2-i\epsilon} = -{\pi^2
T\over R^2}+{\rm subleading}\,.
\eeq
This gives
\beq
f_{2b}=iT\, {GM_1M_2\over R} {G(M_1+M_2)\over 2R}\,.
\eeq
The sum of $f_{2a}$ and $f_{2b}$ gives rise to the post-Newtonian correction
to the potential.

Next, we consider the quantum corrections. First, the polarization contributes
\beq
P_{00\alpha\beta}\Pi^{\alpha\beta\gamma\delta}P_{\gamma\delta
00}={\kappa^2\over 32\pi^2} q^4 \left({43\over 120}\right)
(-\ln(-q^2))\,.
\eeq
The vacuum polarization Fig.\ 3a receives contribution from both the
graviton loop and the Fadeev-Popov-Feynman-Mandelstam ghost loop; this
contribution has been computed more than two decades ago by Capper et
al\cite{cap} and Duff \cite{duff}, and by 't Hooft and Veltman
\cite{velt}. 't Hooft and Veltman obtain
general results that can be transcribed to the long range potential
that we are computing. The result can be written in terms of
coordinate invariant quantities quadratic in the Ricci tensor and the
curvature scalar. Donoghue \cite{donoghue} has quoted their result in
the form given below. We have not checked the complete calculation of
the vacuum polarization from first principles.
Recalling that the propagator is corrected as
\beq
{i\over q^2} P_{0000}+{i\over
q^2}P_{00\alpha\beta}i\Pi^{\alpha\beta\gamma\delta}{i\over q^2}
P_{\gamma\delta 00}=i\left(-{1\over 2 \vec{q}^2}+{G\over \pi}{43\over
120}\ln\vec{q}^2\right)\,,
\eeq
the vacuum polarization contribution to the quantum corrections
follows.

Fig.~3b is calculated easiest in configuration space,
\beq
f_{3b}={1\over 2}\left({i\kappa^2\over 4}\right)^2 M_1M_2
\left(-{1\over 8\pi^2}\right)^2\int dt_1 dt_2 {1\over
((t_1-t_2)^2-R^2-i\epsilon)^2}\,.
\eeq
One notes that the square of the propagator is proportional to the
derivative of a single propagator with respect to $R$.
Recalling (2.19) and (2.20), we obtain
\beq
f_{3b}=iT {GM_1M_2\over R} {G\over 4\pi R^2}\,.
\eeq

The quantum correction of the graph in Fig.~3c involves the
same triple graviton vertex and calculates to
\beq
f_{3c}=-T{\kappa^4M_1M_2\over 32}\int {d^3\vec{p}\over
(2\pi)^3}e^{-i\vec{p}\cdot\vec{R}}\int {d^4 q \over (2\pi)^4}
\left.\left({1\over q^2(p+q)^2}+ 2{q_0^2\over
p^2q^2(p+q)^2}\right)\right|_{p^0=0}\,.
\eeq
Since
\begin{eqnarray}
\int {d^4 q \over (2\pi)^4}\left.{{q_0}^2\over
p^2q^2(p+q)^2}\right|_{p^0=0} &=& -{p^2\over 12}\int {d^4 q \over (2\pi)^4}
\left.{1\over q^2(p+q)^2}\right|_{p^0=0}\\
&=&-{p^2\over 12} \left(-{i\over 16\pi^2}\ln(-p^2)+p{\rm -independent}
\right) \,,
\end{eqnarray}
as can be shown in dimensional
regularization, one obtains the advertised result, after discarding
unimportant short-range divergences. It is useful to note that
\beq
\int {d^3\vec{p}\over
(2\pi)^3}e^{-i\vec{p}\cdot\vec{R}} \ln \vec{p}^2 = -{1\over 2\pi
R^3}\,.
\eeq

Notice that other surviving zero mass non-gravitational contributions
i.e.\ photons, massless neutrinos etc.\ will modify the results of
this appendix, namely the vacuum polarization loop Eq.\ (A.10), and
produce additive contributions as expected. Massive fields will
of course produce forces of finite range which do not survive at large
distances.

Notice also that the quantum corrections have a sign that increases
the attractive gravitational interaction between massive sources as
the separation between constituents is decreased. This sign has a
physical origin. Namely the long range forces can be traced
to real parts of amplitudes with positive definite absorptive parts
coming from physical intermediate states in the crossed channel which
is the source of the logarithms; See Eq.~(A.15), for
example. Unphysical polarizations in such intermediate states are
cancelled by ghost loops as required by unitarity. Hence, positivity
and the associated Schwarz inequality guarantee the
definite sign of the three contributions of Fig.~3. While this is
similar to quantum electrodynamics whose vacuum is dielectric in
nature with dielectric constant greater than one; it is totally at
variance with QCD. In QCD contributions purely real in origin reverse
the sign of coupling constant renormalization and produce the
inevitable asymptotic freedom and its associated paramagnetic vacuum.

\end{document}